\documentclass[5p,times,twocolumn,a4paper,showpacs,showkeys,amsmath,amssymb,superscriptaddress,nofootinbib]{elsarticle}

\usepackage{lineno,hyperref}
\usepackage{comment}
\usepackage{upgreek}

\usepackage[whole]{bxcjkjatype} 
\usepackage{color} 

\modulolinenumbers[5]

\newcommand{\figref}[1]{Fig.~\ref{#1}}
\newcommand{\tabref}[1]{Table~\ref{#1}}
\newcommand{\eqref}[1]{Eq.~\ref{#1}}
\newcommand{\secref}[1]{Sec.~\ref{#1}}

\journal{Journal of \LaTeX\ Templates}

\begin{document}

\title{Prototype study of $0.1\%\,X_0$ and $\mathrm{MHz/cm^2}$ tolerant Resistive Plate Chamber with Diamond-Like Carbon electrodes}

\author[icepp]{Kei Ieki}
\author[u-tokyo]{Weiyuan Li}
\author[kobe-u]{Atsuhiko Ochi}
\author[u-tokyo]{Rina Onda}
\author[icepp]{Wataru Ootani}
\author[u-tokyo]{Atsushi Oya\corref{mycorrespondingauthor}}
\cortext[mycorrespondingauthor]{Corresponding author}
\ead{atsushi@icepp.s.u-tokyo.ac.jp}
\author[kobe-u]{Masato Takahashi}
\author[u-tokyo]{Kensuke Yamamoto}

\address[u-tokyo]{Department of Physics, The University of Tokyo, Bunkyo-ku, Tokyo 113-0033, Japan}
\address[icepp]{International Center for Elementary Particle Physics, The University of Tokyo, Bunkyo-ku, Tokyo 113-0033, Japan}
\address[kobe-u]{Department of Physics, Kobe University, Kobe 657-8501 Japan}

\begin{frontmatter}

\begin{abstract}
A novel Resistive Plate Chamber (RPC) was designed with Diamond-Like Carbon (DLC) electrodes and performance studies were carried out for 384$\,\mathrm{\upmu m}$ gap configuration with a $2\,\mathrm{cm}\times2\,\mathrm{cm}$ prototype.
The use of thin films coated with DLC enables an ultra-low mass design of $< 0.1\%\,X_0$ with up to a four-layer configuration.
At the same time, 42\% MIP efficiency, and 180\,ps timing resolution per layer were achieved in a measurement performed under a $1\,\mathrm{MHz/cm^2}$ non-MIP charged particle beam.
In addition, we propose a further improved design for a $20\,\mathrm{cm}$-scale detector that can achieve 90\% four-layer efficiency in an even higher $4\,\mathrm{MHz/cm^2}$ beam.
In this paper, we describe the detector design, present the results of performance measurements, and characterize the rate capability of the DLC-based RPCs with a performance projection for an improved design.
\end{abstract}


\end{frontmatter}


\section{Introduction}
In the MEG~II experiment searching for the $\mu\to e\gamma$ decay, the reduction of background photons plays an important role in having a high sensitivity.
One of the main sources of such photons is the radiative muon decay (RMD, $\mu\to e\nu\bar{\nu}\gamma$), which can be reduced in the analysis by detecting the coincident positrons typically with $< 5\,\mathrm{MeV}$.
With the magnetic field of the spectrometer, about 90\% acceptance can be covered for the RMD positrons with 20\,cm diameter detectors placed at both ends of the MEG~II apparatus, upstream and downstream (Fig.~77 of \cite{MEGIIdesign}).
Though the downstream detector is already installed as presented in \cite{megiicollaboration2023operation}, the upstream part is still under development.

This paper focuses on the development of the upstream detector, which is required to detect the RMD positrons in the high-rate low-momentum muon beam. 
To be operational in the environment where the low-momentum ($28\,\mathrm{MeV}/c$) muon beam penetrates the detector, the material budget is limited below $0.1\%\,X_0$, and the rate capability up to $4\,\mathrm{MHz/cm^2}$ is required.
In addition, $>90\%$ MIP efficiency and $<1\,\mathrm{ns}$ timing resolution are required to identify the RMD photons efficiently.

The detector candidate is a new type of Resistive Plate Chamber (RPC) with electrodes based on Diamond-Like Carbon (DLC, amorphous carbon material with controllable resistivity \cite{ROBERTSON1992185}) sputtered on thin polyimide foils.
This paper presents measured performances of a $2\,\mathrm{cm}\times 2\,\mathrm{cm}$ prototype with DLC.
This design overcame the limitation on the rate capability and the material budget of the conventional glass-based RPCs while taking advantage of good efficiency and timing resolution.

The concept of the detector design is described in \secref{sec:Design} with an explanation of the prototype configuration.
The measured performances with the prototype detector in several different conditions are presented in \secref{sec:PerformanceStudies}, and the results are discussed in \secref{sec:Discussion} with a suggestion on the final detector design.

\section{Design of RPC with DLC coated electrodes}\label{sec:Design}
\begin{figure}[tbp]
   \centering
   \includegraphics[width=\linewidth]{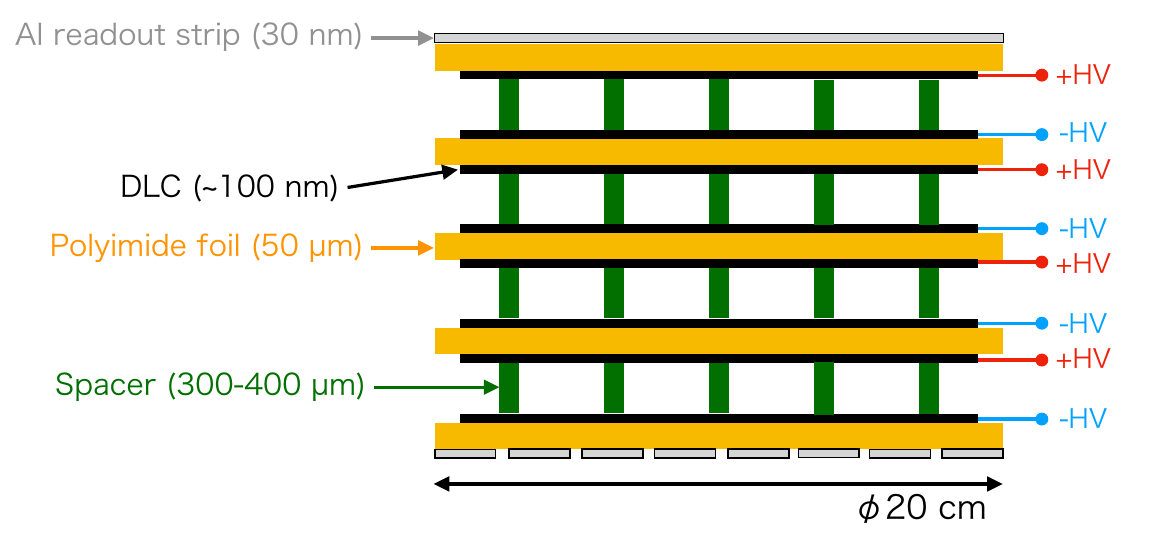}
   \caption{Concept of Resistive Plate Chamber with DLC-based electrodes for the MEG~II experiment. The high voltages are applied independently to each layer. The number of layers is limited because of the $<0.1\%\,X_0$ requirement.}
   \label{fig:DetectorConcept}
\end{figure}

\begin{figure}[tbp]
   \centering
   \includegraphics[width=0.8\linewidth]{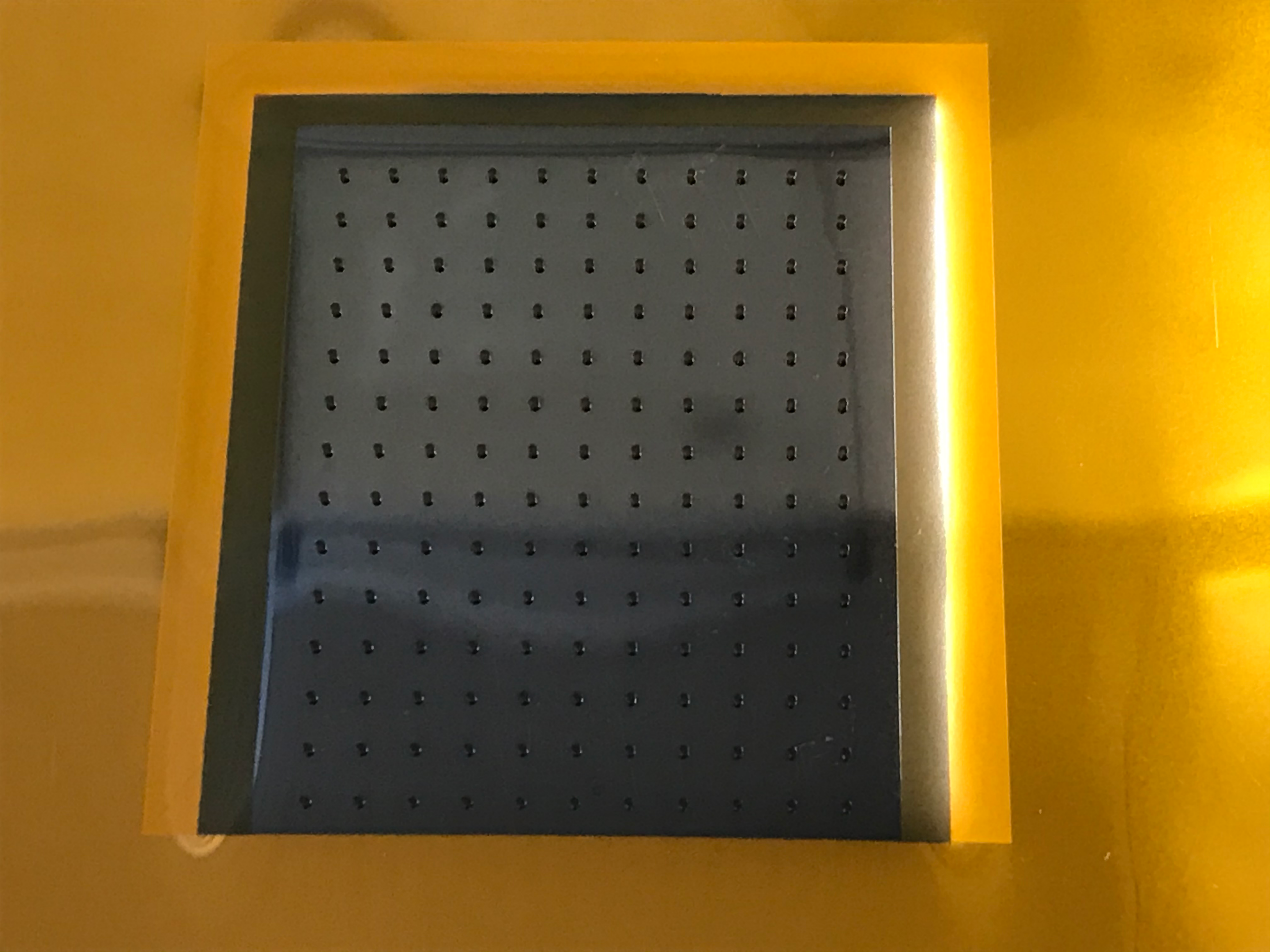}
   \caption{Polyimide film coated with DLC and spacers attached on its surface.}
   \label{fig:DLCFilm}
\end{figure}

The basic structure of the detector is similar to the conventional glass-based RPCs with a uniform electric field between the parallel plate resistive electrodes (\figref{fig:DetectorConcept}).
The difference is that the resistive electrodes consist of 50$\,\mathrm{\upmu m}$ polyimide films coated with DLC (\figref{fig:DLCFilm}).
This offers the advantage of low mass design, and the $<0.1\%\,X_0$ requirement can be fulfilled with up to four layers of configuration.
Another unique characteristic can be found in the high-voltage (HV) supply, particularly in multi-layer configurations.
As polyimide is insulating, the high voltage is applied independently for different layers.
Thus, multi-layer RPCs can be operated with a single-layer voltage, unlike the conventional ones which require $n$ times larger voltage with an $n$-layer configuration.
The high voltage can be supplied by contacting conductive material on the DLC coating.

\subsection*{Design of prototype detector for performance study}
The size of the prototype detector is $2\,\mathrm{cm} \times 2\,\mathrm{cm}$, part of which is read out with a 1\,cm width strip made of aluminized (100\,nm-thick aluminum layer) polyimide.
The signal induced on the strip is amplified by a 38\,dB amplifier and input to a DRS4 waveform digitizer \cite{RITT2004470}.

Cylindrical pillars with 384$\,\mathrm{\upmu m}$ height are used as the spacers for the gas gap, which are attached to the surface of the DLC with photolithography processing. 
They are 400$\,\mathrm{\upmu m}$ in diameter and placed in 2.5\,mm pitch (\figref{fig:DLCFilm}), and thus, only 2\% of the whole area is inactive.

The high voltage is supplied by contacting copper tape to the edge of the DLC with conductive adhesive.
The gas mixture consists of Freon~(R134a), iso-$\mathrm{C_4H_{10}}$ and $\mathrm{SF_6}$.
Two different mixtures were used in the performance evaluation, which is summarized in \tabref{tab:GasMixture} with the applied voltage between the 384$\,\mathrm{\upmu m}$ thick gap.
\begin{table}
 \caption{Gas mixing configurations for the performance measurements.}
 \label{tab:GasMixture}
 \centering
 \begin{tabular}{ccccc}
  \hline
  Type & R134a & iso-$\mathrm{C_4H_{10}}$ & $\mathrm{SF_6}$ & Voltage with 384$\,\mathrm{\upmu m}$ gap\\
  \hline
  (1)  & 94\%  & 5\% & 1\% & 2.5--2.8\,kV\\
  (2)  & 93\%  & 0\% & 7\% & 2.7--3.1\,kV\\
  \hline
 \end{tabular}
\end{table}
The surface resistivity of the DLC is 50--70$\,\mathrm{M\Omega/sq}$ for the anode and 6--9$\,\mathrm{M\Omega/sq}$ for the cathode, where the range of the value indicates non-uniformity of the sputtered DLC resistivity.

\section{Performance studies}\label{sec:PerformanceStudies}
Measurements were performed in different conditions to demonstrate the MIP detection efficiency and the timing resolution in high-rate low-momentum muon beam.
Firstly, the efficiency and timing resolution were measured with low-intensity MIP particles with single and multi-layer configurations, as presented in \secref{sec:SingleLayerMIP} and \secref{sec:MultiLayerMIP}, respectively.
Secondly, the RPC's response to a low-momentum muon beam was measured with the single-layer configuration at low intensity, as presented in \secref{sec:LowRateMuon}.
Finally, the MIP detection performance with the single-layer configuration was measured in a high-rate low-momentum muon beam, as presented in \secref{sec:HighRateMuon}.

\subsection{MIP measurement with single-layer configuration}\label{sec:SingleLayerMIP}
\begin{figure}[tbp]
   \centering
   \includegraphics[width=\linewidth]{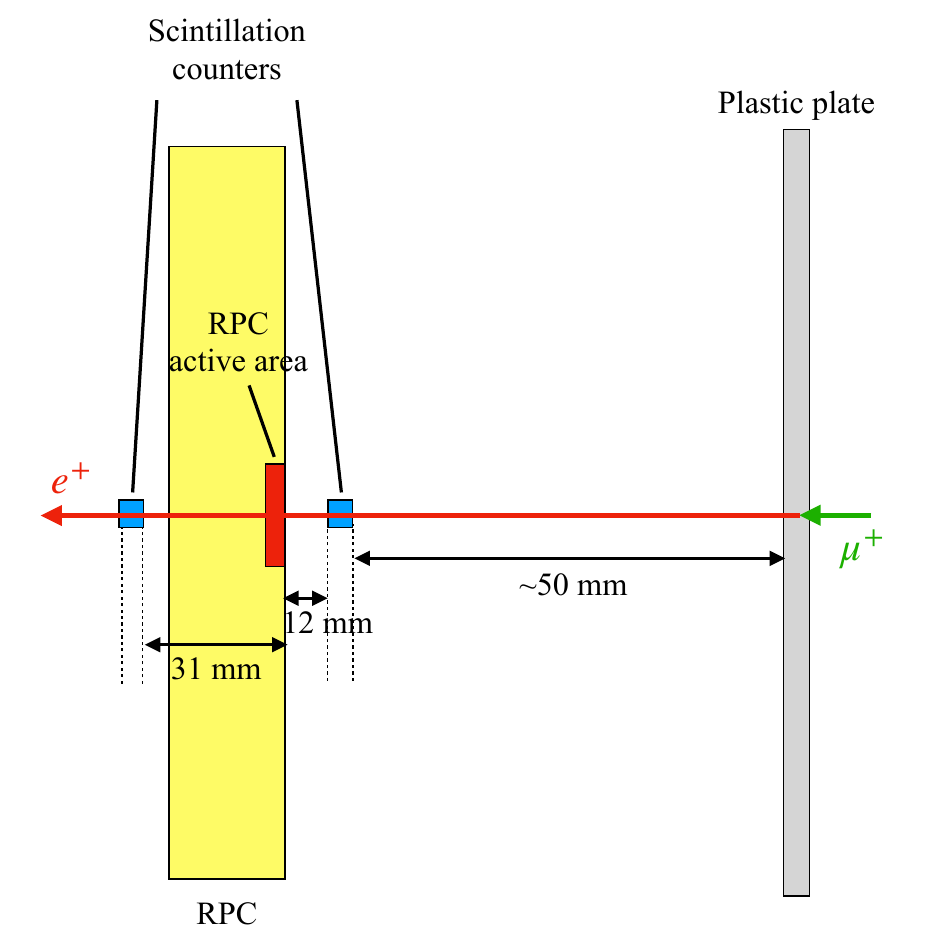}
   \caption{Setup for the test with positrons from muon decays. Muons in the beam were stopped in the plastic plate in front of the detector, and the decay positrons were measured by the RPC. The DAQ was triggered by a coincidence of two scintillation counters.}
   \label{fig:MichelSetup}
\end{figure}

The performance for MIP positrons from muon decays was measured with the configuration shown in \figref{fig:MichelSetup}.
Events were triggered by the coincidence of the two scintillation counters in this measurement, where the low-momentum muon beam was blocked upstream so as not to contaminate the measurement with the beam muon.
This measurement was performed with a single-layer configuration and type(1) gas mixture in \tabref{tab:GasMixture}.

\begin{figure}[tbp]
   \centering
   \includegraphics[width=\linewidth]{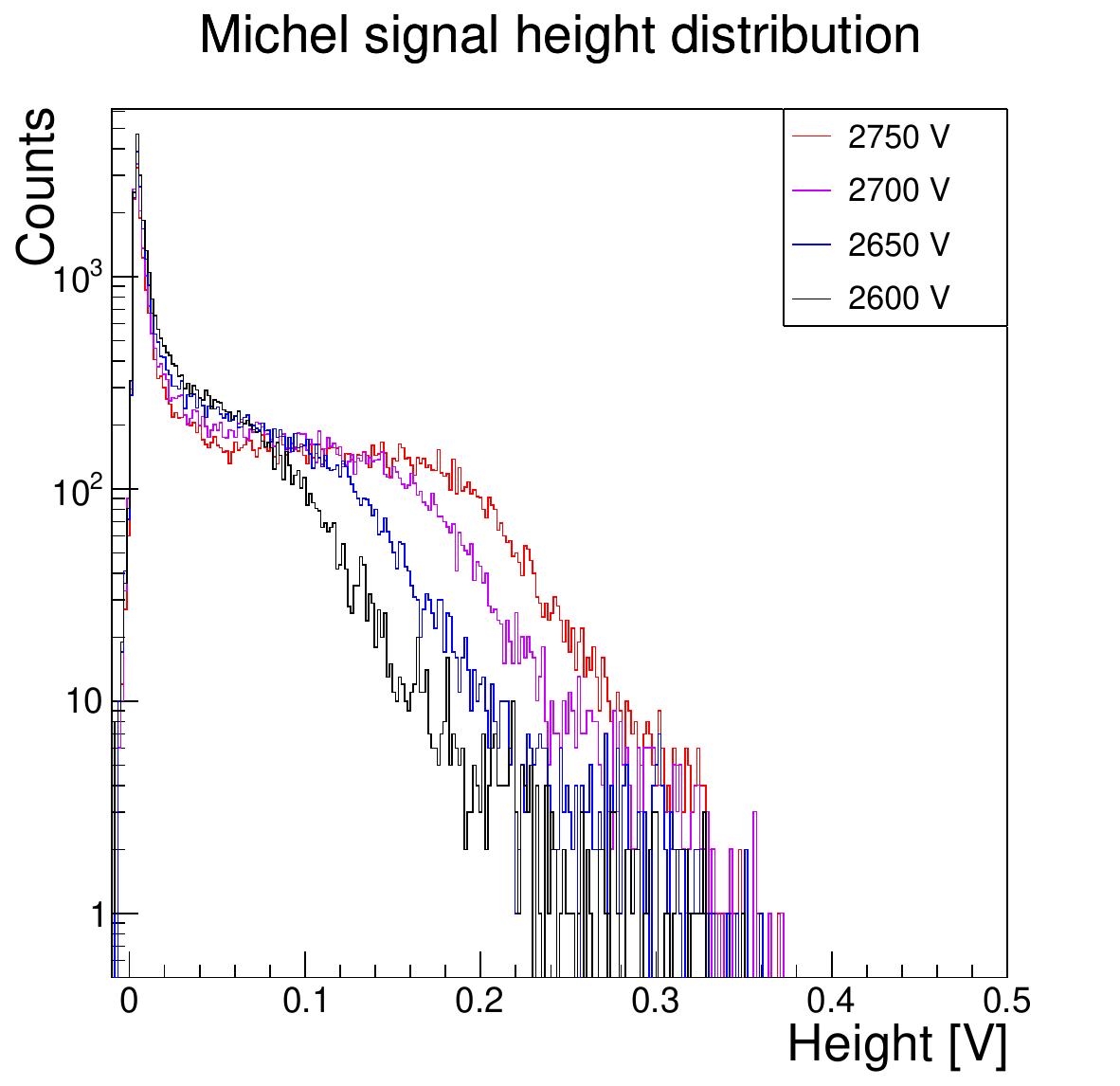}
   \caption{Pulse height spectra of the RPC signal for positrons from muon decays at four different voltages.}
   \label{fig:LowRateMichelSpectra}
\end{figure}

The pulse height spectra for the Michel positron were obtained as shown in \figref{fig:LowRateMichelSpectra}, where we can see a good agreement with \cite{FONTE200217,LIPPMANN200454} in its shape.
When 2.75\,kV is applied between 384$\,\mathrm{\upmu m}$ thick gas gap, the detection efficiency for the single-layer configuration is 55$\pm$3\% with a 10\,mV pulse detection threshold, which corresponds to 12\,fC threshold in the prompt signal charge.

From the timing difference distribution between the RPC and the reference counters, the timing resolution was evaluated to be 170\,ps. 
In a previous study \cite{RIEGLER2003144}, it is pointed out that the timing resolution can be roughly estimated from the effective Townsend coefficient and the drift velocity with the following approximate relation,
\begin{equation}
\Delta t \sim \frac{1.28}{(\alpha-\eta)\cdot v_{\mathrm{drift}}},
\end{equation}
where $\alpha$, $\eta$, and $v_{\mathrm{drift}}$ are the Townsend coefficient, the attachment coefficient, and the drift velocity, respectively.
As a comparison with our measurement, the relevant parameters were simulated with Magboltz \cite{Magboltz}, and we got $\alpha-\eta = 470\,\mathrm{cm^{-1}}$ and $v_{\mathrm{drift}}=0.015\,\mathrm{cm/ns}$.
This gives 180\,ps expectation to the timing resolution, in agreement with the measured one.

\subsection{Demonstration of efficiency improvement with multi-layer configuration}\label{sec:MultiLayerMIP}
\begin{figure}[tbp]
   \centering
   \includegraphics[width=0.7\linewidth]{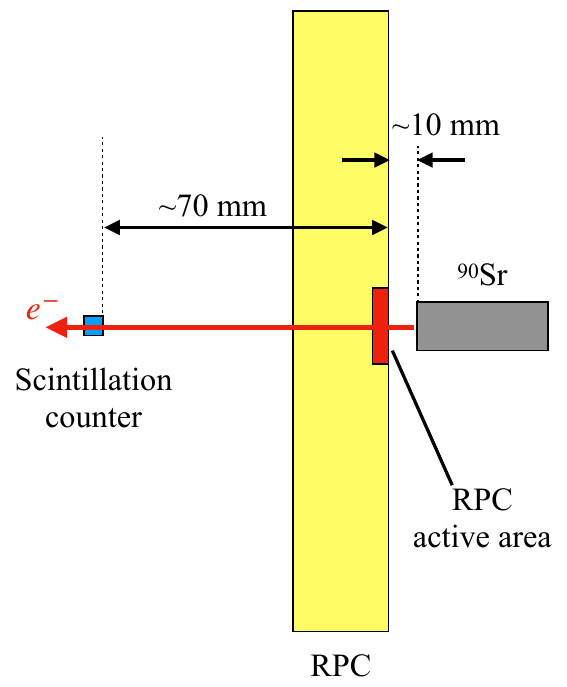}
   \caption{Setup for the test with beta-rays from a $\mathrm{{}^{90}Sr}$ source. The RPC was irradiated at a $\sim100\,\mathrm{kHz/cm^2}$ rate and the DAQ was triggered by the scintillation counter behind the RPC.}
   \label{fig:Sr90Setup}
\end{figure}

The performance for beta-rays from ${}^{90}\mathrm{Sr}$ was measured as shown in \figref{fig:Sr90Setup}, with the type(2) gas mixture in \tabref{tab:GasMixture}.
The measurements, aiming to demonstrate the efficiency improvement in multi-layer configurations, were carried out with the number of layers ranging from one to four. 

\begin{figure}[tbp]
   \centering
   \includegraphics[width=\linewidth]{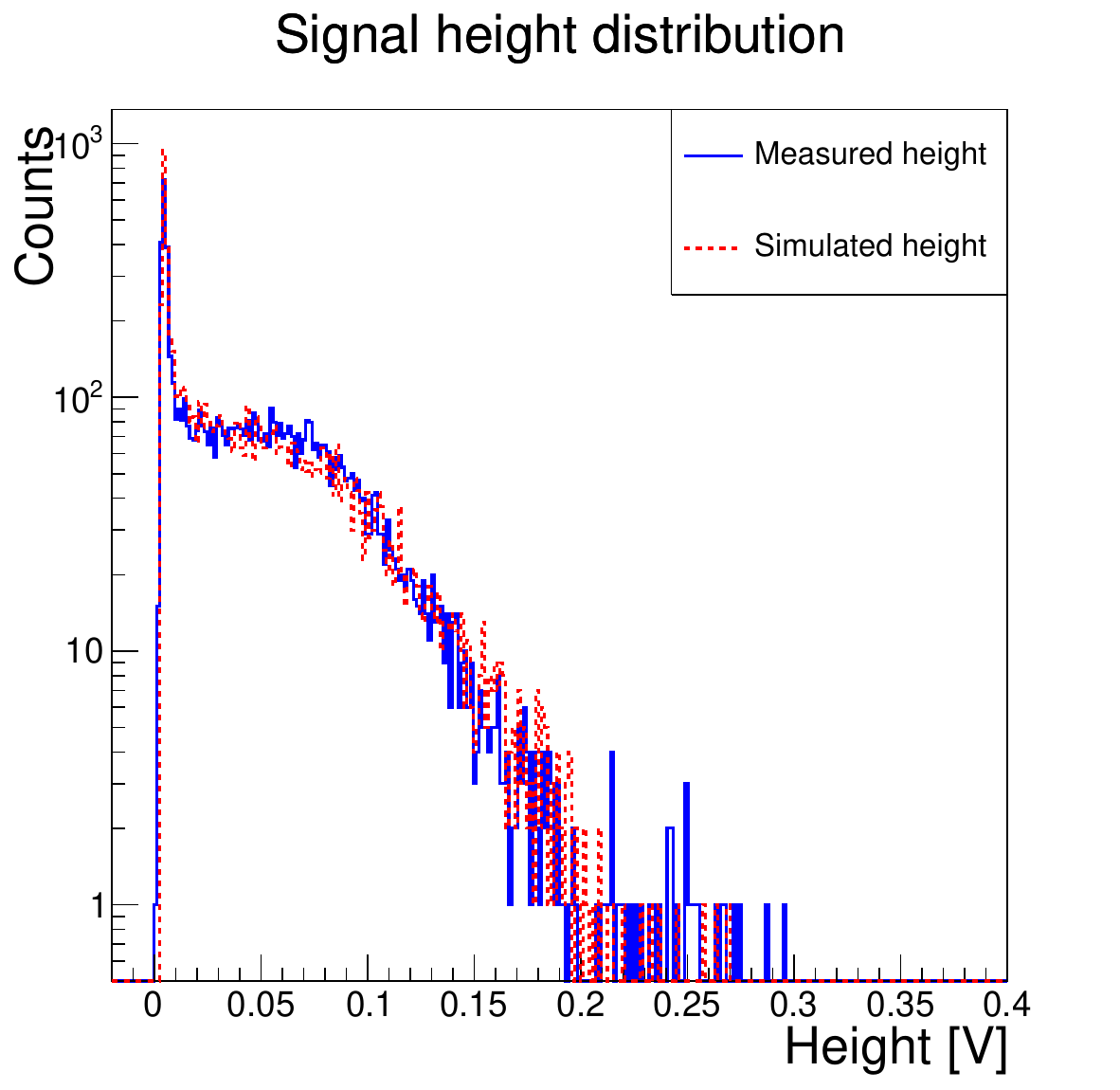}
   \caption{Pulse height spectra obtained for two-layer configuration with beta-ray from $\mathrm{{}^{90}Sr}$ (blue solid line) and pulse height spectra predicted from the single-layer spectra (red dotted line). }
   \label{fig:TwoLayerSpectraSr}
\end{figure}

The blue solid line in \figref{fig:TwoLayerSpectraSr} shows the pulse height spectra for the two-layer configuration. 
The RPC's signal can be understood with Ramo-Shockley's formula \cite{1686997,RIEGLER2002258},
\begin{equation}\label{eq:Ramo'sTh}
i = \sum_{\mathrm{electrons}} e\vec{E}_{\mathrm{w}}\cdot \vec{v}_{\mathrm{drift}},
\end{equation}
where $i$ is the induced current on the readout strip, $\vec{E}_{\mathrm{w}}$ is the weighting field and $\vec{v}_{\mathrm{drift}}$ is the drift velocity of electrons, and the sum is taken over all the electrons inside the gaps.
Thus, the pulse height spectra for multi-layer configuration can be predicted by summing up the contribution of electrons from different layers and then scaling the result with the difference in $\vec{E}_{\mathrm{w}}$ strength.
Overlaid in \figref{fig:TwoLayerSpectraSr} as the red dotted line is the pulse height spectra predicted from the single-layer distribution and \eqref{eq:Ramo'sTh}, and we can see a good agreement between the measurement and the prediction.

\begin{figure}[tbp]
   \centering
   \includegraphics[width=\linewidth]{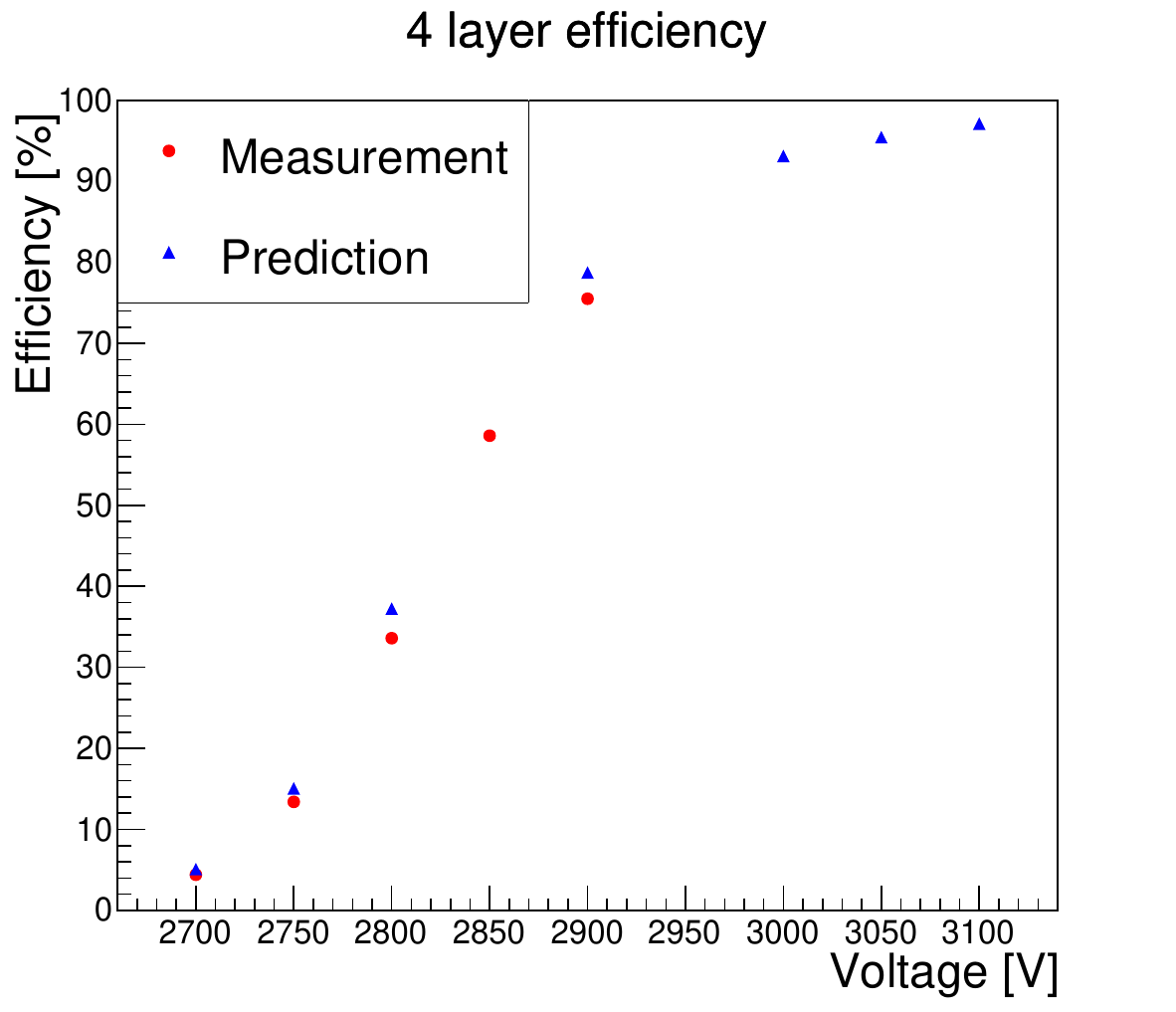}
   \caption{Efficiency with the four-layer configuration of measurement (red circle markers) and prediction (blue triangle markers) from the single-layer results and \eqref{eq:MultiEff}.}
   \label{fig:FourLayerEfficiency}
\end{figure}

From the above consideration, the multi-layer efficiency can be roughly predicted from the single-layer one with 
\begin{equation}\label{eq:MultiEff}
1-\epsilon_n = (1-\epsilon_1)^n,
\end{equation}
where $\epsilon_n$ is the efficiency for $n$-layer configuration \cite{RIEGLER2003144}.
The measured four-layer efficiency is shown in \figref{fig:FourLayerEfficiency} as a function of the applied voltage. 
The results agreed with the predictions from single-layer results and \eqref{eq:MultiEff}, 
though this measurement was performed only up to 2.9\,kV due to discharge, which is understood to originate from the mechanical non-flatness of the electrodes.
These results also suggest that the target single-layer efficiency is $> 40\%$ to achieve 90\% overall efficiency with the proposed four-layer configuration.

\subsection{Low-momentum muon measurement}\label{sec:LowRateMuon}
\begin{figure}[tbp]
   \centering
   \includegraphics[width=\linewidth]{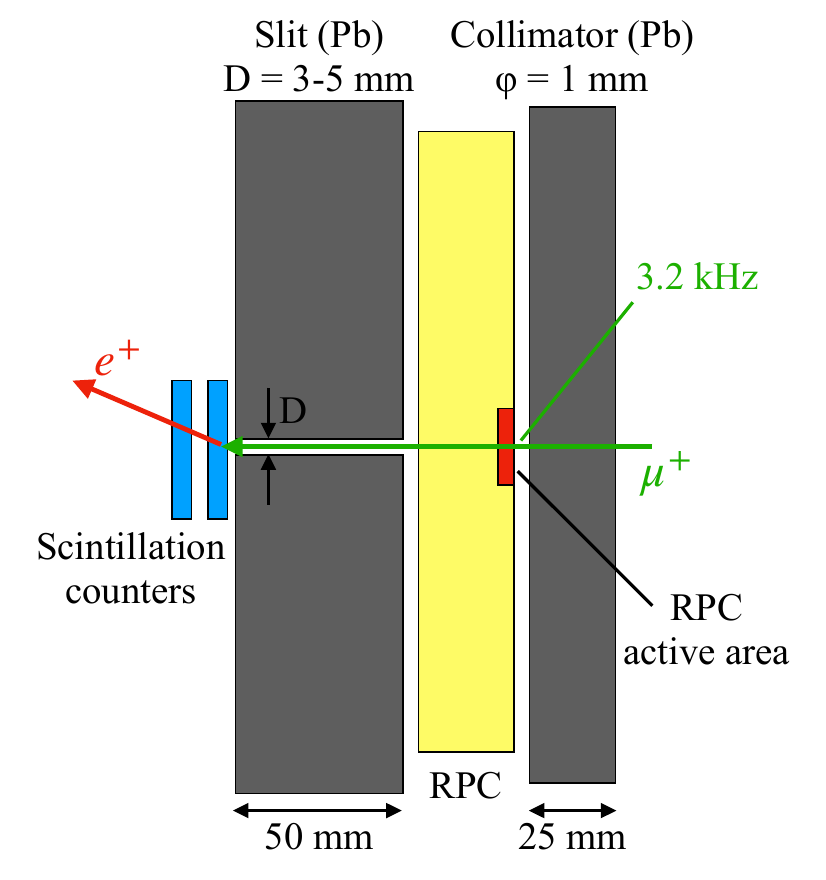}
   \caption{Setup for the test with low-momentum muons. The muon beam was collimated down to $\mathcal{O}(\mathrm{kHz/cm^2})$ by the collimator in front of the RPC. The DAQ was triggered by a delayed coincidence of two scintillation counters. The lead slit behind the RPC was put in order to reduce fake triggers from background radiations in this measurement.}
   \label{fig:LowRateSetup}
\end{figure}

The response to the low-momentum muon is measured at the $\pi E5$ beam line of Paul Scherrer Institute with the single-layer RPC filled with the type(1) gas mixture of \tabref{tab:GasMixture}.
To decouple the effect of the rate capability, this measurement was conducted with the beam collimated down to $\mathcal{O}(\mathrm{kHz/cm^2})$, as shown in \figref{fig:LowRateSetup}.
This rate was confirmed to be low enough with cross-checking measurements at different beam rates.
The trigger was issued by a delayed coincidence of muon and a subsequent positron, which was judged by a primary hit only to the first counter followed by a hit to both counters within 100--450\,ns after the primary hit.

\begin{figure}[tbp]
   \centering
   \includegraphics[width=\linewidth]{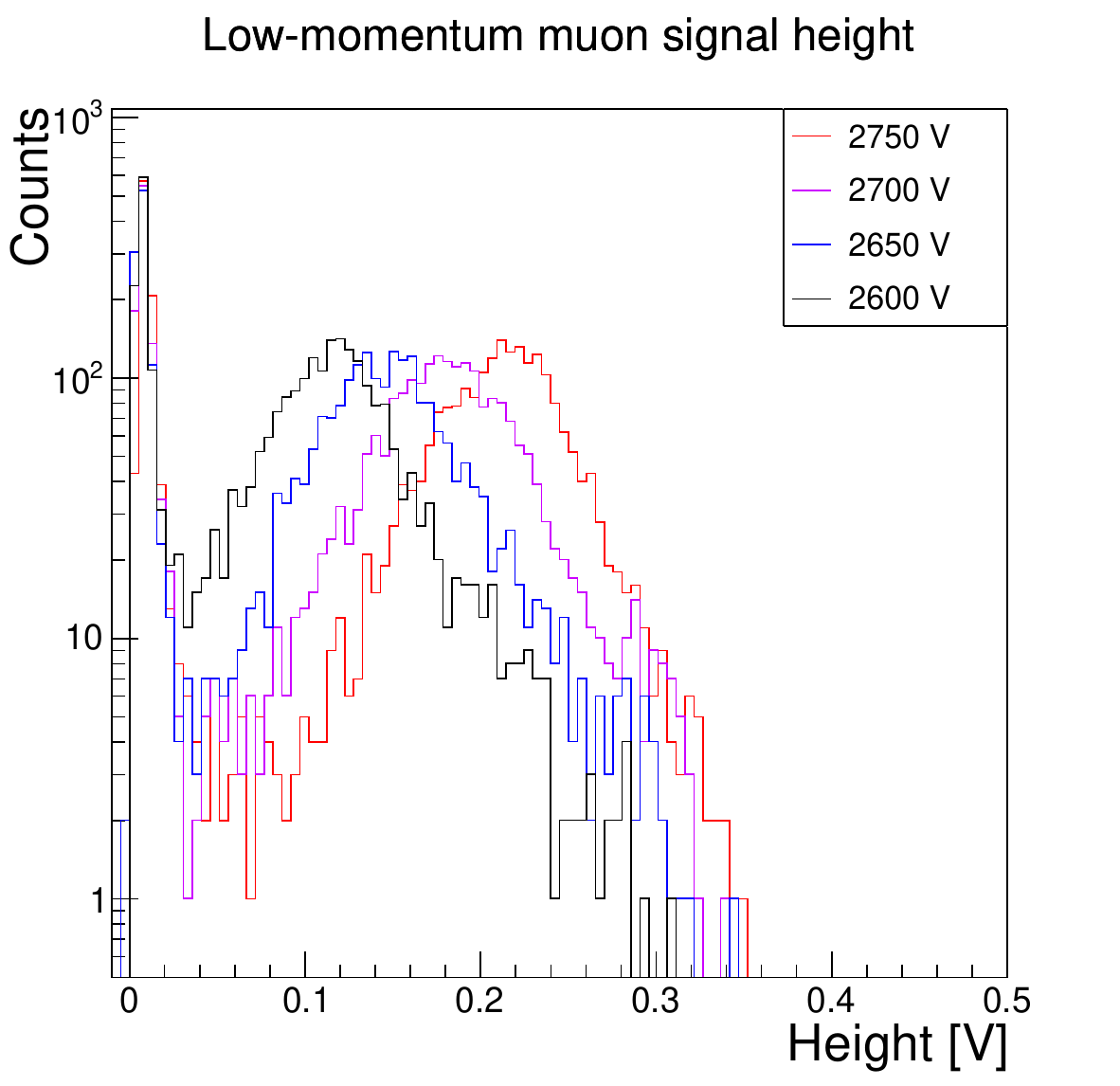}
   \caption{Pulse height spectra obtained for low-momentum muons at different operation voltages. The main peak around 0.1--0.2\,V comes from RPC signals for low-momentum muons. The peak around 0\,V is understood to be fake triggered events, not a real inefficiency of the RPC.}
   \label{fig:LowRateMuonSpectra}
\end{figure}

The pulse height spectra for all the triggered events are shown in \figref{fig:LowRateMuonSpectra}, with the applied voltage between 2.6--2.75\,kV.
The main peak around 0.1--0.2\,V comes from the muon events, whereas the peak around $0\,\mathrm{V}$ is understood as fake triggered events coming from background radiation.
This background radiation is understood to be photons radiated from the decay positrons interacting inside the collimator.
Though the exact single-layer efficiency for low-momentum muon was not successfully measured with this background, it is at least $>75\%$ with an applied voltage of 2.75\,kV, evaluated from the fraction of the peak events.

\figref{fig:LowRateMuonSpectra} also suggests that the pulse height spectra for low-momentum muon stays at the same level as that for MIP, even though the energy deposit by low-momentum muon is an order of magnitude larger.
This can be explained by the space charge effect, which saturates the avalanche growth when the gain reaches $\mathcal{O}(10^7)$ \cite{LIPPMANN200454}.
As will be discussed in detail in \secref{sec:HighRateMuon}, the rate capability is determined by the gain reduction due to voltage drop, which originates from the signal current flowing on the electrodes.
Thus, this saturation effect is advantageous from a rate capability viewpoint.

\subsection{Performance in high-rate low-momentum muon beam}\label{sec:HighRateMuon}
\begin{figure}[tbp]
   \centering
   \includegraphics[width=\linewidth]{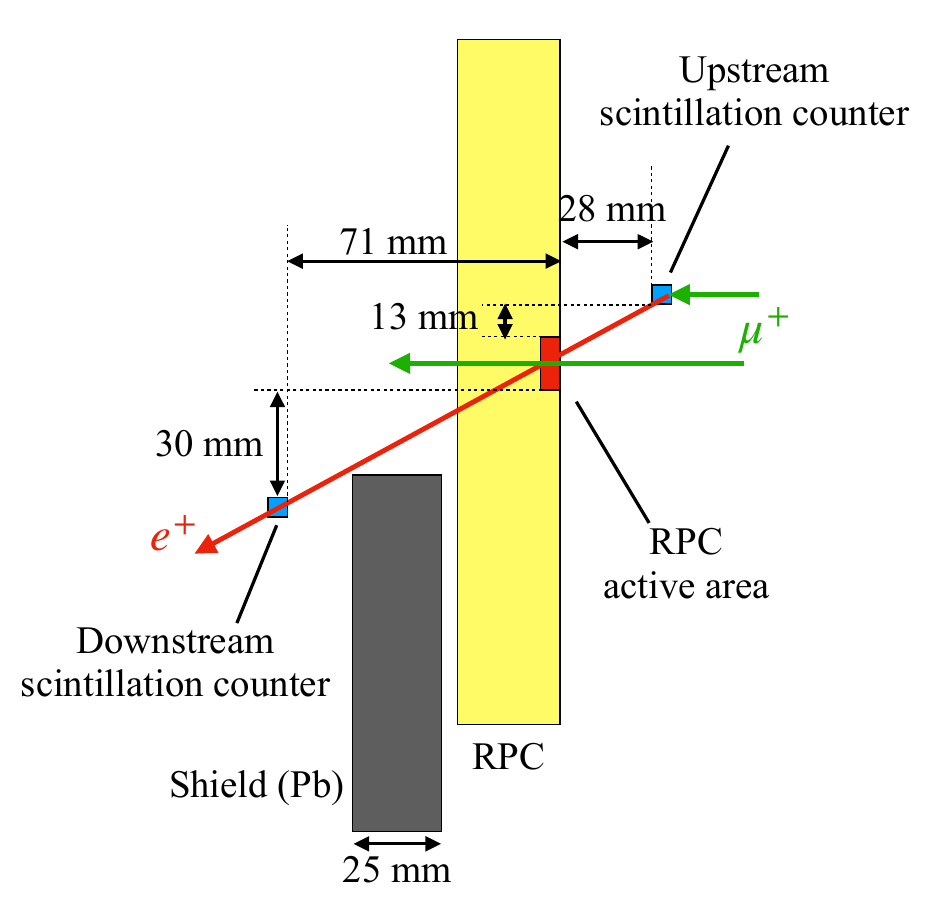}
   \caption{Setup for MIP detection test in high-rate low-momentum muon beam. The RPC was irradiated with a rate of 1\,$\mathrm{MHz/cm^2}$. The DAQ was triggered by a positron generating a coincidence of two scintillation counters. The lead shield behind the RPC is to reduce the hit rate of the backward scintillation counter, which thus reduces the accidental trigger.}
   \label{fig:HighRateSetup}
\end{figure}

\begin{figure}[tbp]
   \centering
   \includegraphics[width=\linewidth]{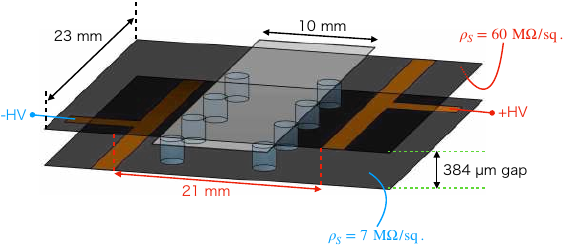}
   \caption{The alignment of RPC detector components. Two HV supply lines, one on the anode and the other on the cathode, were placed with a 21\,cm pitch. The aluminum readout strip was located in the middle of the HV lines. The beam center was aligned to the center of the readout strip.}
   \label{fig:HighRateHVSupply}
\end{figure}

The detection efficiency for MIP positrons in a high-intensity low-momentum muon beam was measured with the setup shown in \figref{fig:HighRateSetup}.
The beam intensity at the center was 1\,$\mathrm{MHz/cm^2}$ with the spread of $(\sigma_x,\sigma_y) = (13\,\mathrm{mm},23\,\mathrm{mm})$.
The RPC was filled with the gas mixture of type(1) in \tabref{tab:GasMixture} and the applied voltage was always 2.75\,kV.
The detailed structure of RPC components is shown in \figref{fig:HighRateHVSupply}, where the center of the muon beam was aligned to the center of the readout strip with a $\sim$ mm precision.
The trigger was issued by positrons from muon decays by taking the coincidence of two counters, which were aligned for triggered positrons (from decays of muon stopped in the upstream one) to cross the active region of the RPC.

\begin{figure}[tbp]
   \centering
   \includegraphics[width=\linewidth]{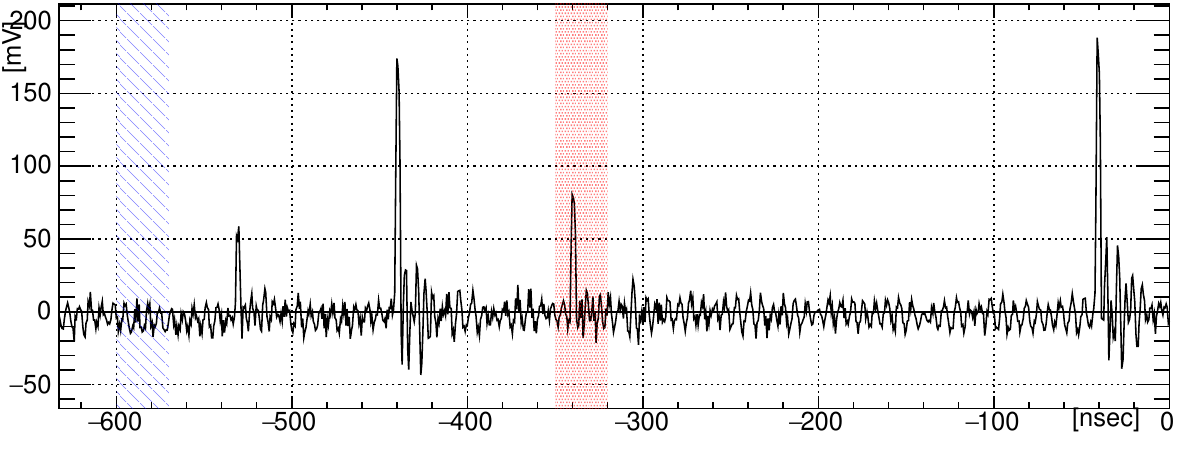}
   \caption{An example waveform in the measurement with the high-rate muon beam. The red-hatched (blue-shaded) region is the analysis window for trigger timing (off timing). The observed pulse in the triggered time window is likely a positron from a muon decay, and the other three pulses are from beam muons.}
   \label{fig:WaveformHighRate}
\end{figure}

A typical waveform is shown in \figref{fig:WaveformHighRate}, where the red hatched region is a window for trigger timing analysis, and a positron from a muon decay is expected inside.
On the other hand, the blue-shaded region is a window for off-timing analysis, in which accidental beam muons are expected.

\begin{figure}[tbp]
   \centering
   \includegraphics[width=\linewidth]{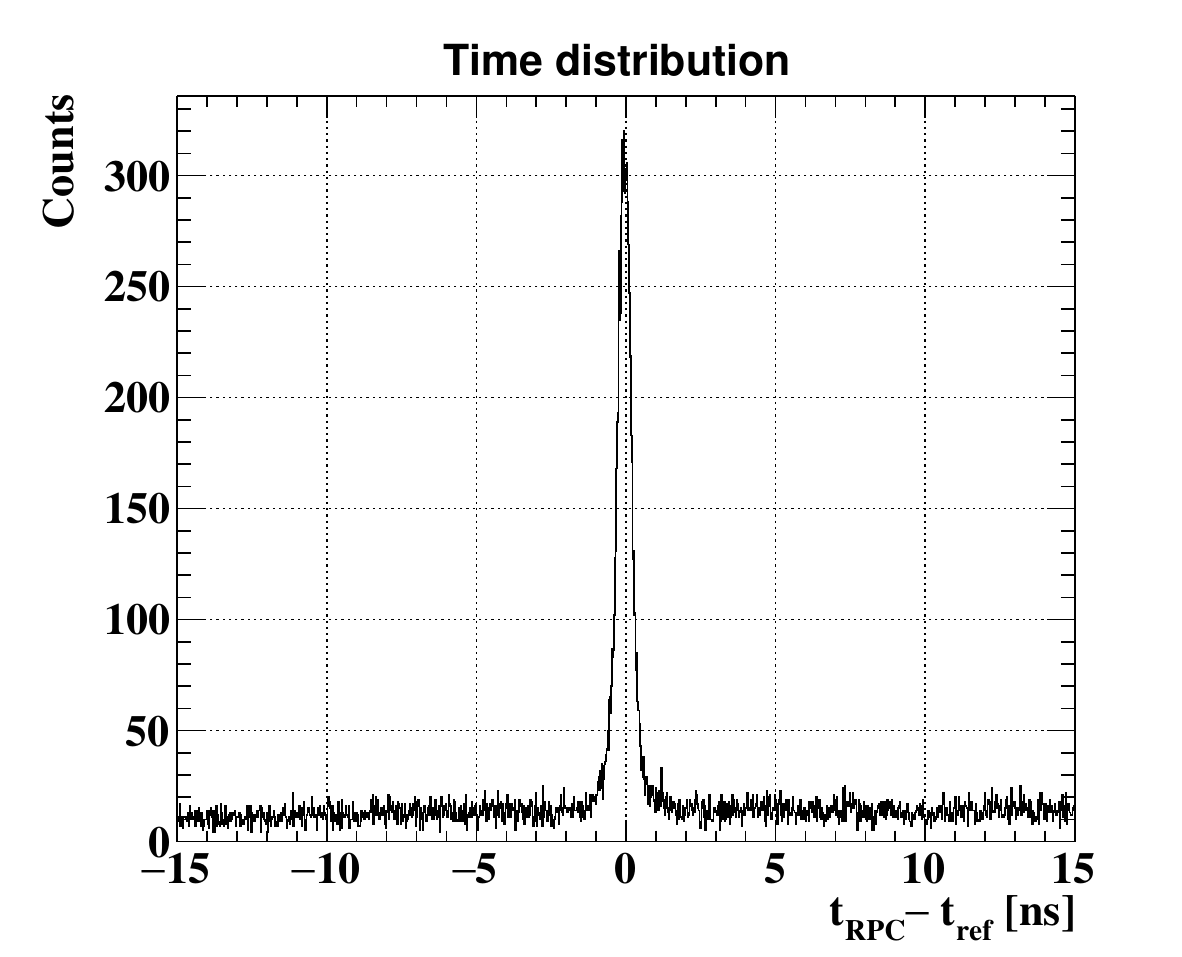}
   \caption{Time difference between the RPC and the trigger counter in the high-rate measurement. The peak comes from detected positrons and the baseline comes from accidental muon hits.}
   \label{fig:HighRateTimeDist}
\end{figure}

\begin{figure}[tbp]
   \centering
   \includegraphics[width=\linewidth]{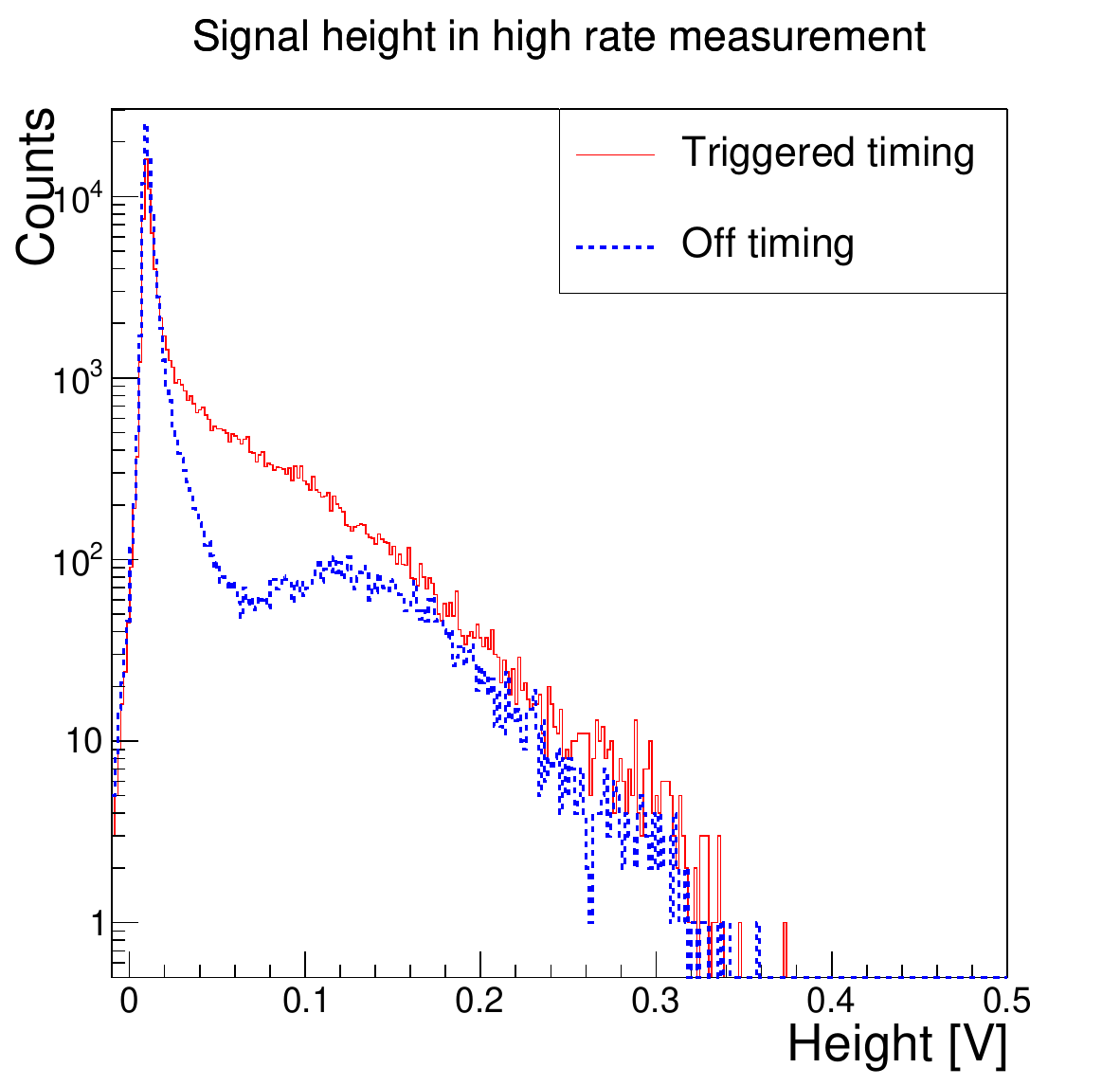}
   \caption{Pulse height spectra for on-timing (red solid line) and off-timing (blue dotted line) in high-rate low-momentum muon beam}
   \label{fig:HighRateSpectra}
\end{figure}

The signal time distribution referenced by the trigger counter is shown in \figref{fig:HighRateTimeDist}, where we found 180\,ps timing resolution, which is comparable to the low-rate resolution of 170\,ps in \secref{sec:SingleLayerMIP}.
The muon hit rate estimated from the sideband region agreed with the expectation from the rate and profile of the beam.
The signal height distribution in the on-timing and off-timing analysis window is depicted in \figref{fig:HighRateSpectra} as the red solid line and the blue dotted line, respectively.
The on-timing signal height distribution agreed with the 2.6--2.65\,kV distribution in \figref{fig:LowRateMichelSpectra}, which is 100--150\,V lower than the nominal applied voltage.
This is due to the voltage drop effect in a high-rate measurement, which determines the rate capability of RPCs.
The off-timing height distribution has a bump around 0.12\,V, which can be understood to originate from beam muons.
The bump peak position also agreed with the 2.6--2.65\,kV distribution shown in \figref{fig:LowRateMuonSpectra}.
The positron efficiency was also evaluated by counting the events inside the peak in \figref{fig:HighRateTimeDist}, and we got 42\% as expected from the observed voltage drop effect. 
The observed efficiency reduction can be mitigated by improving the design, as will be discussed in \secref{sec:Discussion}.

In high-rate irradiation, we expect a gain degradation due to a voltage drop coming from the high signal current flowing on the high resistivity electrodes \cite{Abbrescia_2016,Aielli_2016,CARBONI2003135}.
With signal current on the DLC surface, the voltage drop follows
\begin{equation}\label{eq:VoltageDrop}
\left(\frac{\partial^2}{\partial x^2}+\frac{\partial^2}{\partial y^2}\right) V_{\mathrm{drop}}(x,y) = Q_{\mathrm{mean}} \cdot f(x,y) \cdot \rho,
\end{equation}
where $V_{\mathrm{drop}}(x,y)$ is the voltage drop at position $(x,y)$, $Q_{\mathrm{mean}}$ is the average avalanche charge for each incoming particle, $f(x,y)$ is the position-dependent beam rate per unit area, and $\rho$ is the surface resistivity of DLC.
The average avalanche charge in this measurement was evaluated to be $Q_{\mathrm{mean}}\sim 2.3\pm0.2\,\mathrm{pC}$ from the current observed at the HV supply divided by the expected total hit rate inside the RPC's sensitive volume.
The boundary condition of \eqref{eq:VoltageDrop} is determined by the geometry of the HV supply.
With the inputs above, the simulated voltage drop in the readout region is 100--150\,V depending on the position, well in agreement with the observation.

\section{Discussion}\label{sec:Discussion}
\subsection{Further improvement of rate capability}
\begin{figure}[tbp]
   \centering
   \includegraphics[width=\linewidth]{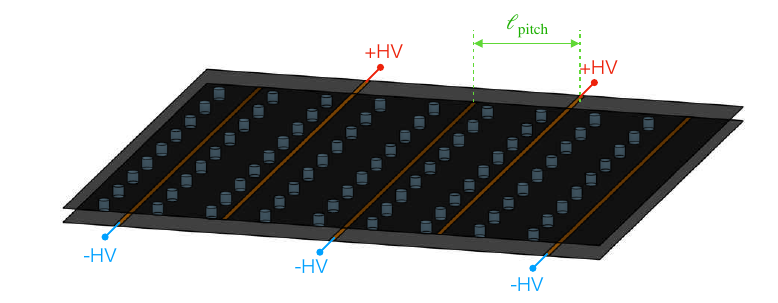}
   \caption{The design concept of strip HV supply. Strip-shaped HV supply lines are put on the anode and cathode surfaces with an $\ell_\mathrm{pitch}$ pitch.}
   \label{fig:stripHVSupply}
\end{figure}

As is implied in \eqref{eq:VoltageDrop}, the voltage drop can be further suppressed with a smaller distance to the HV supply and lower surface resistivity.
To limit the distance to HV supply with large area RPCs, a segmented HV supply is necessary.
One candidate idea is to insert conductive strips into the DLC area, as shown in \figref{fig:stripHVSupply}.
With the strip pitch denoted as $\ell_{\mathrm{pitch}}$, the voltage drop roughly scales as $V_{\mathrm{drop}} \propto \ell_{\mathrm{pitch}}^2$.
At the same time, the voltage drop can be suppressed with a lower DLC resistivity with the linear dependence, $V_{\mathrm{drop}} \propto \rho$.
The limitation to $\ell_{\mathrm{pitch}}$ is that the conductive strip region becomes inactive for signal detection, 
and the limitation to $\rho$ is that the detector's stability gets worse with a lower resistivity with a frequent discharge.
Thus, the final design must be optimized with a further understanding of the limitation of the DLC resistivity. 
In our experience, we have already succeeded in a stable operation with the anode resistivity of 20--30$\,\mathrm{M\Omega/sq}$.
Here, we also found that the detector stability is less sensitive to the cathode resistivity, which can be smaller than 1$\,\mathrm{M\Omega/sq}$.

\subsection{Evaluation of ratio between the avalanche charge and the detected charge}
In \secref{sec:HighRateMuon}, the evaluation of the voltage drop effect with \eqref{eq:VoltageDrop} was based on the measured avalanche charge of $Q_{\mathrm{mean}}\sim 2.3\pm0.2\,\mathrm{pC}$ at the effective voltage ($V_{\mathrm{eff}}=V_{\mathrm{applied}}-V_{\mathrm{drop}}$) of 2.6--2.65\,kV.
However, the $Q_{\mathrm{mean}}$ itself depends on the effective voltage. 
Therefore, its full dependencies are essential when one tries to estimate the rate capability in different configurations.

The $Q_{\mathrm{mean}}$ vs $V_\mathrm{eff}$ relation can be extrapolated according to the measured pulse height spectra in \figref{fig:LowRateMuonSpectra}.
This is because of the relation between the fast signal charge and the avalanche charge, which was both measured \cite{Aielli_2016} and simulated \cite{RIEGLER2003144}.
These studies point out that the fast signal charge is approximately proportional to the avalanche charge though the coefficient itself depends on the gas gap.
We thus need to find the ratio and then scale the measured pulse height spectra in \figref{fig:LowRateMuonSpectra}.

The ratio evaluation is based on the measurement with the high-rate beam presented in \secref{sec:HighRateMuon}.
Here, the total avalanche charge for low-momentum muons was evaluated to be $2.3\pm0.2\,\mathrm{pC}$. 
On the other hand, the average of the fast signal charge in this measurement was evaluated from the peak of the observed bump at $130\,\mathrm{mV}$.
The fast charge is thus evaluated to be $160\pm30\,\mathrm{fC}$ from the peak height and the shape of the waveform, where the uncertainty mainly comes from our understanding of the electronics response.
Thus, the ratio of the total charge divided by the fast charge is $14\pm4$.

As a cross-check, both the fast charge and the total charge were evaluated also for quasi-MIP beta-rays from $\mathrm{{}^{90}Sr}$.
As a result, the average avalanche charge for the MIP particle was found to be 1.3$\pm$0.3~pC from the measured current and the hit rate.
At the same time, the average of the fast signal charge for the same radiation was evaluated to be $70\pm20\,\mathrm{fC}$ from the observed signal height spectra.
Thus, the ratio of total charge divided by the fast charge is $19\pm6$, in agreement with the ratio obtained for low-momentum muons.
This also agrees with the results obtained for MIP particles in \cite{Aielli_2016}.

\subsection{Expected performance}
Calculated with $V_{\mathrm{applied}}=2.75\,\mathrm{kV}$, $Q_{\mathrm{mean}}=3\,\mathrm{pC}$, $\ell_{\mathrm{pitch}} = 1\,\mathrm{cm}$ and $\rho = 10\,\mathrm{M\Omega/sq}$, the voltage drop is expected to be $\sim 100\,\mathrm{V}$.
Here, $Q_{\mathrm{mean}}$ at the effective voltage (including the $\sim 100\,\mathrm{V}$ drop) is estimated according to the discussion in the previous section.
From the result of the performance measurement at a low rate presented in \secref{sec:SingleLayerMIP}, 40\% single-layer efficiency is achievable with $V_{\mathrm{drop}}\sim 100\,\mathrm{V}$.
Thus, once we establish the design with $\ell_{\mathrm{pitch}} = 1\,\mathrm{cm}$ and $\rho = 10\,\mathrm{M\Omega/sq}$, the required 90\% overall efficiency can be expected with a four-layer configuration even in a low-momentum muon beam at a $4\,\mathrm{MHz/cm^2}$ rate.
At the same time, when we take into account the difference in the avalanche charge between the MIP particles and low-momentum muons, the rate capability against the MIP particles can be as large as $10\,\mathrm{MHz/cm^2}$ with this configuration.


\section{Conclusion}
A low-mass and high-rate capable detector is necessary for further background suppression in the MEG~II experiment.
As a candidate, we proposed a novel design RPC assembled with DLC-based electrodes instead of glasses in the conventional RPCs.
A prototype $2\,\mathrm{cm}\times2\,\mathrm{cm}$ and $< 0.1\%\,X_0$ detector was designed, and performance measurements were carried out in different conditions.
In the measurement with high-intensity low-momentum muon beam at a rate of $1\,\mathrm{MHz/cm^2}$, we achieved 42\% MIP efficiency and 180\,ps time resolution with a single layer configuration.
We quantitatively explained the efficiency degradation in a high-rate beam, considering the voltage drop effect.
Furthermore, we performed a systematic comparison between the voltage drop effect and the measured signal height spectra at a low rate.
Combining the results above, we obtained all the necessary knowledge to estimate the rate capability with a DLC-based RPC detector.
From the above studies, we proposed a further improved design that can achieve $> 90\%$ efficiency with a four-layer configuration, even in a $4\,\mathrm{MHz/cm^2}$ low-momentum muon beam.
Finally, as a side effect of these studies, we also observed the behavior of RPC's gain with the non-MIP charged particles and found that the signal charge is not proportional to the expected energy deposit due to the strong space charge saturation.

\section*{Acknowledgement}
We are grateful for the technical support and cooperation provided by PSI as the host institute.
We would also like to thank Dr. Angela Papa and Dr. Patrick Schwendimann for their cooperation in setting up the muon beam for our measurements.
This work is supported by JSPS Core-to-Core Program, A. Advanced Research Networks JPJSCCA20180004.

\bibliography{mybibfile}

\begin{thebibliography}{10}
\expandafter\ifx\csname url\endcsname\relax
  \def\url#1{\texttt{#1}}\fi
\expandafter\ifx\csname urlprefix\endcsname\relax\def\urlprefix{URL }\fi
\expandafter\ifx\csname href\endcsname\relax
  \def\href#1#2{#2} \def\path#1{#1}\fi

\bibitem{MEGIIdesign}
{Baldini, A. M.}, et~al.,
  \href{https://doi.org/10.1140/epjc/s10052-018-5845-6}{{The design of the
  MEG~II experiment - MEG~II Collaboration}}, Eur. Phys. J. C 78~(5) (2018)
  380.
\newblock \href {http://dx.doi.org/10.1140/epjc/s10052-018-5845-6}
  {\path{doi:10.1140/epjc/s10052-018-5845-6}}.
\newline\urlprefix\url{https://doi.org/10.1140/epjc/s10052-018-5845-6}

\bibitem{megiicollaboration2023operation}
Afanaciev, et~al., Operation and performance of meg ii detector (2023).
\newblock \href {http://arxiv.org/abs/2310.11902} {\path{arXiv:2310.11902}}.

\bibitem{ROBERTSON1992185}
J.~Robertson,
  \href{https://www.sciencedirect.com/science/article/pii/025789729290001Q}{Properties
  of diamond-like carbon}, Surface and Coatings Technology 50~(3) (1992)
  185--203.
\newblock \href
  {http://dx.doi.org/https://doi.org/10.1016/0257-8972(92)90001-Q}
  {\path{doi:https://doi.org/10.1016/0257-8972(92)90001-Q}}.
\newline\urlprefix\url{https://www.sciencedirect.com/science/article/pii/025789729290001Q}

\bibitem{RITT2004470}
S.~Ritt,
  \href{https://www.sciencedirect.com/science/article/pii/S016890020302922X}{The
  drs chip: cheap waveform digitizing in the ghz range}, Nuclear Instruments
  and Methods in Physics Research Section A: Accelerators, Spectrometers,
  Detectors and Associated Equipment 518~(1) (2004) 470--471, frontier
  Detectors for Frontier Physics: Proceedin.
\newblock \href {http://dx.doi.org/https://doi.org/10.1016/j.nima.2003.11.059}
  {\path{doi:https://doi.org/10.1016/j.nima.2003.11.059}}.
\newline\urlprefix\url{https://www.sciencedirect.com/science/article/pii/S016890020302922X}

\bibitem{FONTE200217}
P.~Fonte, V.~Peskov,
  \href{https://www.sciencedirect.com/science/article/pii/S0168900201019143}{High-resolution
  tof with rpcs}, Nuclear Instruments and Methods in Physics Research Section
  A: Accelerators, Spectrometers, Detectors and Associated Equipment 477~(1)
  (2002) 17--22, 5th Int. Conf. on Position-Sensitive Detectors.
\newblock \href
  {http://dx.doi.org/https://doi.org/10.1016/S0168-9002(01)01914-3}
  {\path{doi:https://doi.org/10.1016/S0168-9002(01)01914-3}}.
\newline\urlprefix\url{https://www.sciencedirect.com/science/article/pii/S0168900201019143}

\bibitem{LIPPMANN200454}
C.~Lippmann, W.~Riegler,
  \href{https://www.sciencedirect.com/science/article/pii/S0168900203026421}{Space
  charge effects in resistive plate chambers}, Nuclear Instruments and Methods
  in Physics Research Section A: Accelerators, Spectrometers, Detectors and
  Associated Equipment 517~(1) (2004) 54--76.
\newblock \href {http://dx.doi.org/https://doi.org/10.1016/j.nima.2003.08.174}
  {\path{doi:https://doi.org/10.1016/j.nima.2003.08.174}}.
\newline\urlprefix\url{https://www.sciencedirect.com/science/article/pii/S0168900203026421}

\bibitem{RIEGLER2003144}
W.~Riegler, C.~Lippmann, R.~Veenhof,
  \href{https://www.sciencedirect.com/science/article/pii/S0168900203003371}{Detector
  physics and simulation of resistive plate chambers}, Nuclear Instruments and
  Methods in Physics Research Section A: Accelerators, Spectrometers, Detectors
  and Associated Equipment 500~(1) (2003) 144--162, nIMA Vol 500.
\newblock \href
  {http://dx.doi.org/https://doi.org/10.1016/S0168-9002(03)00337-1}
  {\path{doi:https://doi.org/10.1016/S0168-9002(03)00337-1}}.
\newline\urlprefix\url{https://www.sciencedirect.com/science/article/pii/S0168900203003371}

\bibitem{Magboltz}
\href{https://magboltz.web.cern.ch/magboltz/}{Magboltz - transport of electrons
  in gas mixtures}.
\newline\urlprefix\url{https://magboltz.web.cern.ch/magboltz/}

\bibitem{1686997}
S.~{Ramo}, Currents induced by electron motion, Proceedings of the IRE 27~(9)
  (1939) 584--585.
\newblock \href {http://dx.doi.org/10.1109/JRPROC.1939.228757}
  {\path{doi:10.1109/JRPROC.1939.228757}}.

\bibitem{RIEGLER2002258}
W.~Riegler,
  \href{https://www.sciencedirect.com/science/article/pii/S0168900202011695}{Induced
  signals in resistive plate chambers}, Nuclear Instruments and Methods in
  Physics Research Section A: Accelerators, Spectrometers, Detectors and
  Associated Equipment 491~(1) (2002) 258--271.
\newblock \href
  {http://dx.doi.org/https://doi.org/10.1016/S0168-9002(02)01169-5}
  {\path{doi:https://doi.org/10.1016/S0168-9002(02)01169-5}}.
\newline\urlprefix\url{https://www.sciencedirect.com/science/article/pii/S0168900202011695}

\bibitem{Abbrescia_2016}
M.~Abbrescia, \href{https://doi.org/10.1088/1748-0221/11/10/c10001}{Improving
  rate capability of resistive plate chambers}, Journal of Instrumentation
  11~(10) (2016) C10001--C10001.
\newblock \href {http://dx.doi.org/10.1088/1748-0221/11/10/c10001}
  {\path{doi:10.1088/1748-0221/11/10/c10001}}.
\newline\urlprefix\url{https://doi.org/10.1088/1748-0221/11/10/c10001}

\bibitem{Aielli_2016}
G.~Aielli, P.~Camarri, R.~Cardarelli, A.~D. Ciaccio, L.~D. Stante, R.~Iuppa,
  B.~Liberti, L.~Paolozzi, E.~Pastori, R.~Santonico, M.~Toppi,
  \href{https://doi.org/10.1088/1748-0221/11/07/p07014}{Improving the {RPC}
  rate capability}, Journal of Instrumentation 11~(07) (2016) P07014--P07014.
\newblock \href {http://dx.doi.org/10.1088/1748-0221/11/07/p07014}
  {\path{doi:10.1088/1748-0221/11/07/p07014}}.
\newline\urlprefix\url{https://doi.org/10.1088/1748-0221/11/07/p07014}

\bibitem{CARBONI2003135}
G.~Carboni, G.~Collazuol, S.~{De Capua}, D.~Domenici, G.~Ganis, R.~Messi,
  G.~Passaleva, E.~Santovetti, M.~Veltri,
  \href{https://www.sciencedirect.com/science/article/pii/S016890020202082X}{A
  model for rpc detectors operating at high rate}, Nuclear Instruments and
  Methods in Physics Research Section A: Accelerators, Spectrometers, Detectors
  and Associated Equipment 498~(1) (2003) 135--142.
\newblock \href
  {http://dx.doi.org/https://doi.org/10.1016/S0168-9002(02)02082-X}
  {\path{doi:https://doi.org/10.1016/S0168-9002(02)02082-X}}.
\newline\urlprefix\url{https://www.sciencedirect.com/science/article/pii/S016890020202082X}

\end{thebibliography}

\end{document}